\documentclass {epl}

\usepackage {epsfig}
\usepackage {amsmath,amssymb}

\title{Wall effects on granular heap stability}
\author{S. Courrech du Pont\inst{1}
\and P. Gondret\inst{1}
\and B. Perrin\inst{2}
\and M. Rabaud\inst{1}}
\institute{
    \inst{1} F.A.S.T. Universit\'es Paris 6 \& 11 et C.N.R.S. (U.M.R. 
    7608),\\ b{\^a}t. 502, campus universitaire, 91405 
    Orsay cedex, France\\
    \inst{2} L.P.M.C. Universit\'es Paris 6 \& 7, E.N.S. et C.N.R.S. 
    (U.M.R. 8551),\\ 24 rue Lhomond, 75005 Paris, France\\}


\begin{document}
\maketitle
\begin{abstract}
We investigate the effects of lateral walls on the angle of movement
and on the angle of repose of a granular pile.  Our experimental
results for beads immersed in water are similar to previous results
obtained in air and to recent numerical simulations. All of these
results, showing an increase of pile angles with a decreasing gap
width, are explained by a model based on the redirection of stresses
through the granular media.  Two regimes are observed depending on the
bead diameter.  For large beads, the range of wall effects corresponds
to a constant number of beads whereas it corresponds to a constant
characteristic length for small beads as they aggregate via van der
Waals forces.
\end{abstract}                               
A characteristic of a sand pile is that it forms a non-zero angle to the horizontal. Two angles can be defined for 
a heap of granular matter: the angle of repose $\theta_{\rm{r}}$, under which no flow can occur, and, a few degrees larger, 
the maximum angle of stability $\theta_{\rm{m}}$ first noticed by Bagnold \cite{mouv}. This angle, also called the angle of movement, 
is the one at which an avalanche spontaneously occurs at the surface of the pile, making the slope angle relax 
to the angle of repose. Between these two angles is a region of bistability 
as the heap can be static (``solid state") 
or flowing (``liquid state"). The values of these two angles have been known for long to depend on many parameters, 
namely the shape, the roughness, the size distribution and the packing fraction of grains, as well as the packing 
history \cite{divers}. Humidity, by introducing cohesion through capillary bridges between grains, is known to strongly 
increase the stability of a heap \cite{cohesion}.
The presence of close lateral walls, by changing the boundary conditions, also 
increases the stability of a heap, as both angles $\theta_{\rm{m}}$ 
and $\theta_{\rm{r}}$ increase when the gap width between the confining walls 
decreases \cite{liu}.  
This is often explained qualitatively by the presence of particle arches between the walls \cite{gras,bolt,zhou}. Particle 
arches or divergence of force networks lead to many remarkable effects. The saturation of the pressure at 
the bottom of containers, known as the Janssen effect \cite{jans}, makes hourglasses flow at constant speed. In silos, 
arch formation may lead to a complete jamming of the flow, with potential damages for industry.
Most of laboratory model experiments on granular physics are led using experimental set-up where granular media is confined.
Lateral wall effects on pile angles have recently been studied both experimentally 
\cite{gras,bolt,zhou} and numerically \cite{zhou}.
Grasselli and Herrmann \cite{gras}
measured the angle of repose for dry glass beads of different diameter 
$d$ ($0.1 \leq d \leq 0.5 \, \rm{mm}$), by slowly filling 
from one side a rectangular glass cell of gap width $b$ in the range $1 < 
b < 10 \, \rm{mm}$. Boltenhagen ~\cite{bolt} measured the 
angle of movement of dry millimetric glass beads ($1 \leq d \leq 4 \, \rm{mm}$) by tilting a Plexiglas rectangular 
cell ($8 < b/d < 100$). More recently, Zhou and co-workers ~\cite{zhou} have investigated the angle of 
repose both experimentally and numerically by discharging a rectangular 
box ($4 < b/d < 24$) filled 
with dry monodisperse glass spheres ($1 \leq d \leq 10 \, \rm{mm}$). 
In all these experiments the wall effects are only controlled by the 
smallest dimension (gap width $b$), {\it{i.e.}} other cell dimensions do not play any role.
All these authors \cite{gras,bolt,zhou} observed that the pile angles 
$\theta_{\rm{r}}$ and $\theta_{\rm{m}}$ decrease with an increasing gap
width towards constant values, and fitted their results by the empirical 
exponential law
\begin{equation}
\label{eq1}
\theta=\theta_{\infty}[1+\alpha\exp(-b/b^\ast)],
\end{equation}
with three fitting parameters: $\theta_{\infty}$ is the angle value when $b$ tends towards infinity, 
$b^\ast$ is the characteristic length scale of wall effect and $\alpha$ is a parameter 
of order 1 \cite{gras,bolt,zhou}. This description brings to wonder which parameters govern the characteristic 
length scale $b^\ast$ of wall effects. 
Boltenhagen found for millimetric beads that the characteristic number 
of beads $n^\ast$, defined as $n^\ast=b^\ast/d$ , decreases 
slightly with increasing bead diameters ~\cite{bolt}. The results of Zhou 
{\it{et al.}} \cite{zhou} can be interpreted as a constant $n^\ast$ 
value ($\sim \, 6$), independent of the particle size. 
At the opposite, Grasselli and Herrmann found for submillimetric 
beads that $n^\ast$ strongly increases as the bead diameter decreases, leading to a constant value of $b^\ast$ ($\sim \, 9 \, \rm{mm}$) 
and evoked 
the cohesion due to humidity as a possible explanation for this unexpected 
behaviour ~\cite{gras}.\\
We have experimentally studied for a large range of bead diameters the lateral wall 
effects on both the angle of movement and the angle of repose of a packing of glass spheres.
To eliminate the effect of possible capillary bridges between beads and to reduce the electrostatic 
forces, glass beads are totally immersed in water. We also propose a physical modelling of 
the wall influence on pile angles that is based on the Janssen effect 
~\cite{jans}, {\it{i.e.}} on the pressure 
saturation that occurs with depth in confined granular media.
\begin{figure}[h!]
    
    \begin{minipage}{0.47\linewidth}
	\centerline{\epsfig{file=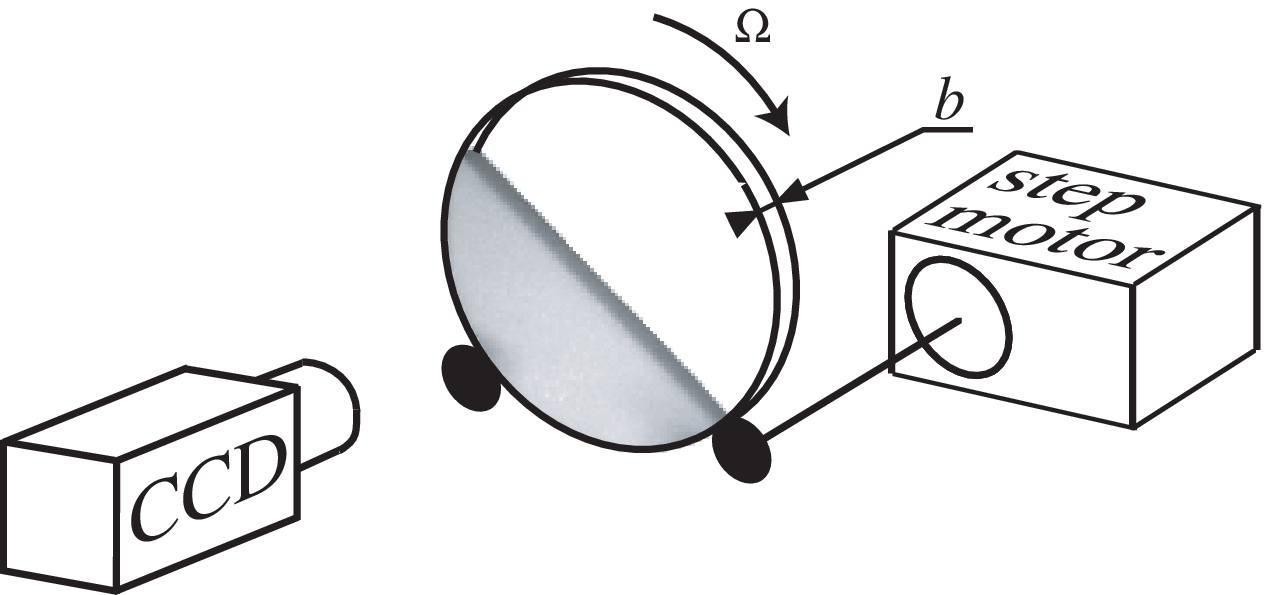,width=0.8\linewidth}}
    \end{minipage}
    \hfill
    \begin{minipage}{0.47\linewidth}
	\centerline{\epsfig{file=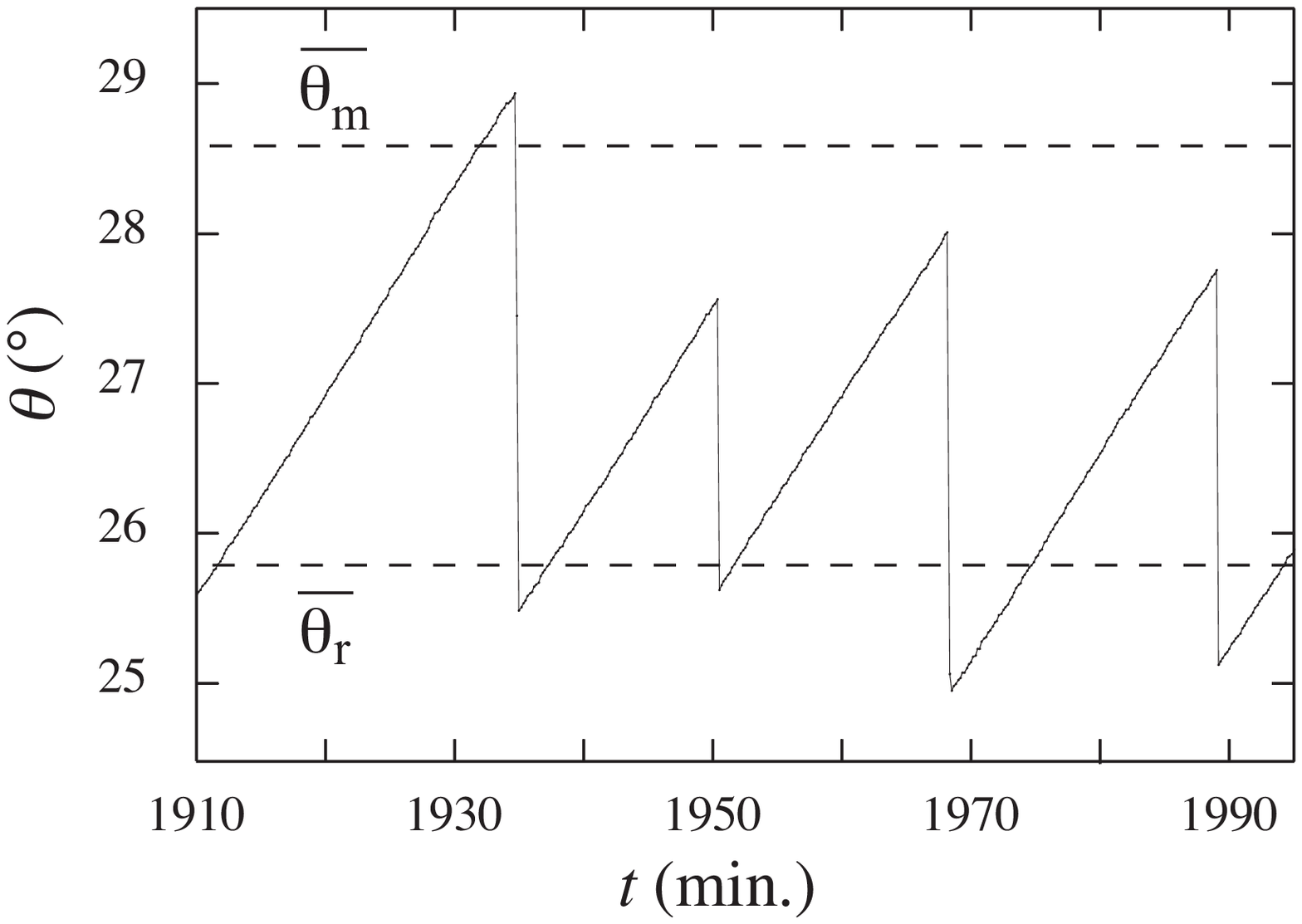,width=0.8\linewidth}}
    \end{minipage}
    \caption{Sketch of the rotating drum experiment.}
    \label{fig1}
    \caption{Time evolution of the slope angle $\theta$ for glass beads of diameter $d = 1.85 \, \rm{mm}$ 
    totally immersed in water, in a cylinder of gap width $b = 15.5 \, \rm{mm}$ rotating at the angular 
    velocity $\Omega = 4.10^{-3}\, \rm{rpm}$. From the whole experiment, $\overline{\theta_{\rm{m}}}= 28.6 \pm 0.8^{\circ}$ and 
    $\overline{\theta_{\rm{r}}}= 25.8 \pm 0.5^{\circ}$ (dashed lines).}
    \label{fig2}
\end{figure}

A sketch of our experimental set-up is displayed in fig. 1. It consists of a rotating cylinder
of inner diameter $D = 17 \, \rm{cm}$, half-filled with rather monodisperse sieved glass 
beads of mean diameter $d$ (with a dispersion of 10\%).
The granular pile is totally immersed in water and confined between two parallel glass or Plexiglas
endwalls separated by a gap whose width $b$ can be varied thanks to rubber wedges. The cylinder lies 
on two horizontal parallel axes, one of which is driven by a micro-step motor followed by a reducer, 
so that the cylinder smoothly turns at a constant angular velocity 
$\Omega$. The angular velocity $\Omega$ is chosen 
low enough to be in the intermittent regime of avalanches 
~\cite{intermit}, so that the typical time between two avalanches 
is much larger than the typical avalanche duration. With a CCD video camera set in the laboratory frame and 
aligned with the cylinder axis, images are taken at regular time intervals, then analysed in order to track 
the pile interface. This interface is found to be plane, thus well characterised 
by its mean slope angle $\theta$. 
Any change of $0.01^{\circ}$ for the mean slope angle can be detected.\\
A typical time evolution of the mean slope angle $\theta(t)$ is displayed in fig. 2. 
The slope angle increases linearly with time at the rate $\Omega$, as the pile is in solid 
rotation with the drum, up to the maximum angle of stability 
$\theta_{\rm{m}}$. Then it quickly relaxes through 
a surface avalanche down to the angle of repose $\theta_{\rm{r}}$. Over all the duration of one experiment 
(more than one hundred events), we calculate the mean values 
$\overline{\theta_{\rm{m}}}$ and $\overline{\theta_{\rm{r}}}$. Henceforth, we focus 
on these mean values, and relieve the mean bar notations. Note that the fluctuations around these 
mean values are found to be smaller for the angle of repose than for the angle of movement. In the 
present experiments we have made the diameter of the glass beads vary in 
the range $0.1 \, \rm{mm} < d \leq 3 \, \rm{mm}$, 
and the gap width of the cylinder in the range $1 \leq b \leq 60 \, \rm{mm}$ and 
$2 < b/d < 100$. The cylinder aspect ratio $b/D$ remains small ($6 \, 10^{-3} 
\lesssim b/D \lesssim 3 \, 10^{-1}$) in order to avoid any diameter effects. 
This explains that 
observed avalanches concern the whole width of the pile and leave the 
interface flat, contrary to large $b/D$ studies \cite{fauve}.
\begin{figure}[h!]
    \begin{minipage}[b]{0.47\linewidth}
	\centerline{\epsfig{file=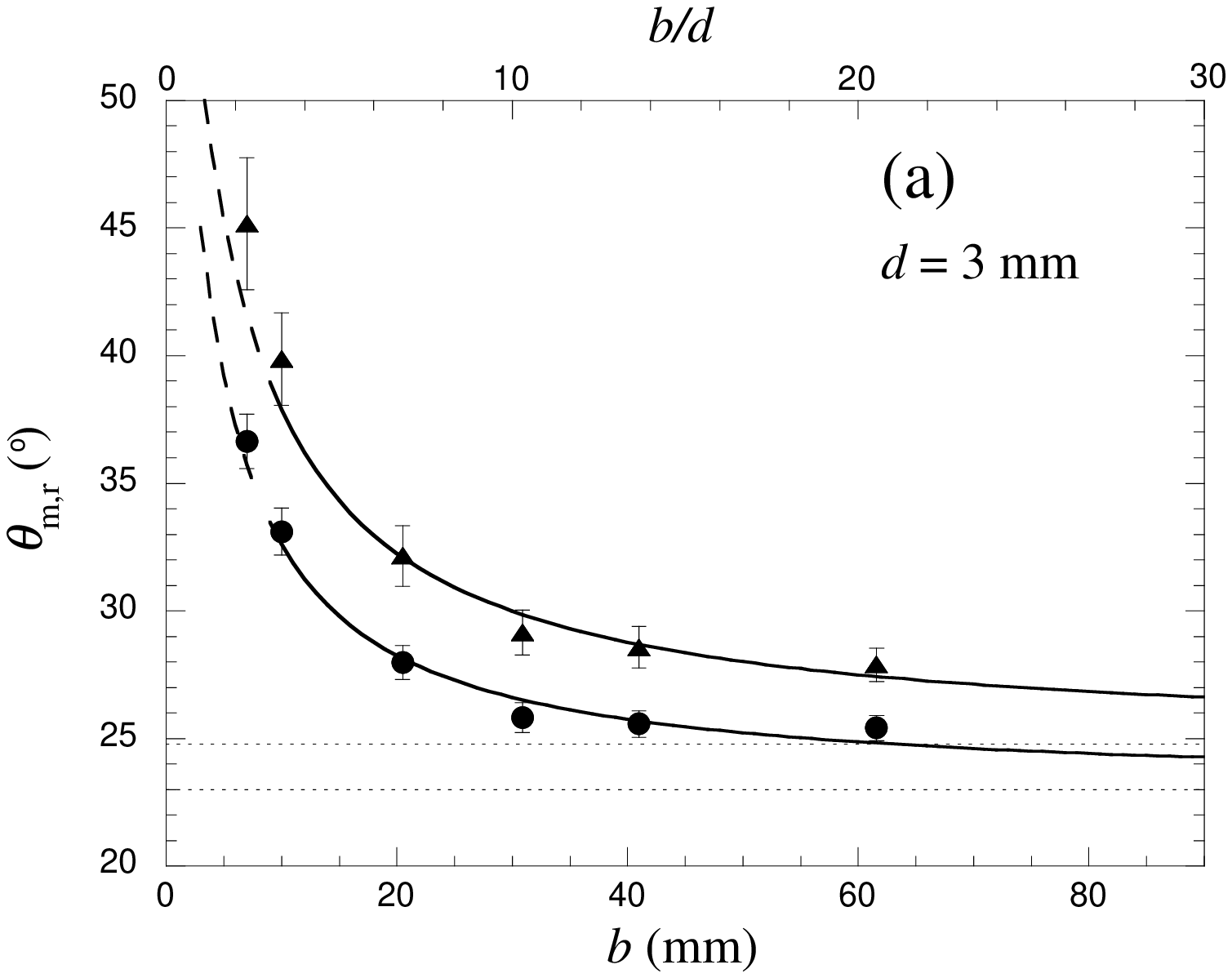,width=0.9\linewidth}}
    \end{minipage}
 \hfill
    \begin{minipage}[b]{0.47\linewidth}
	\centerline{\epsfig{file=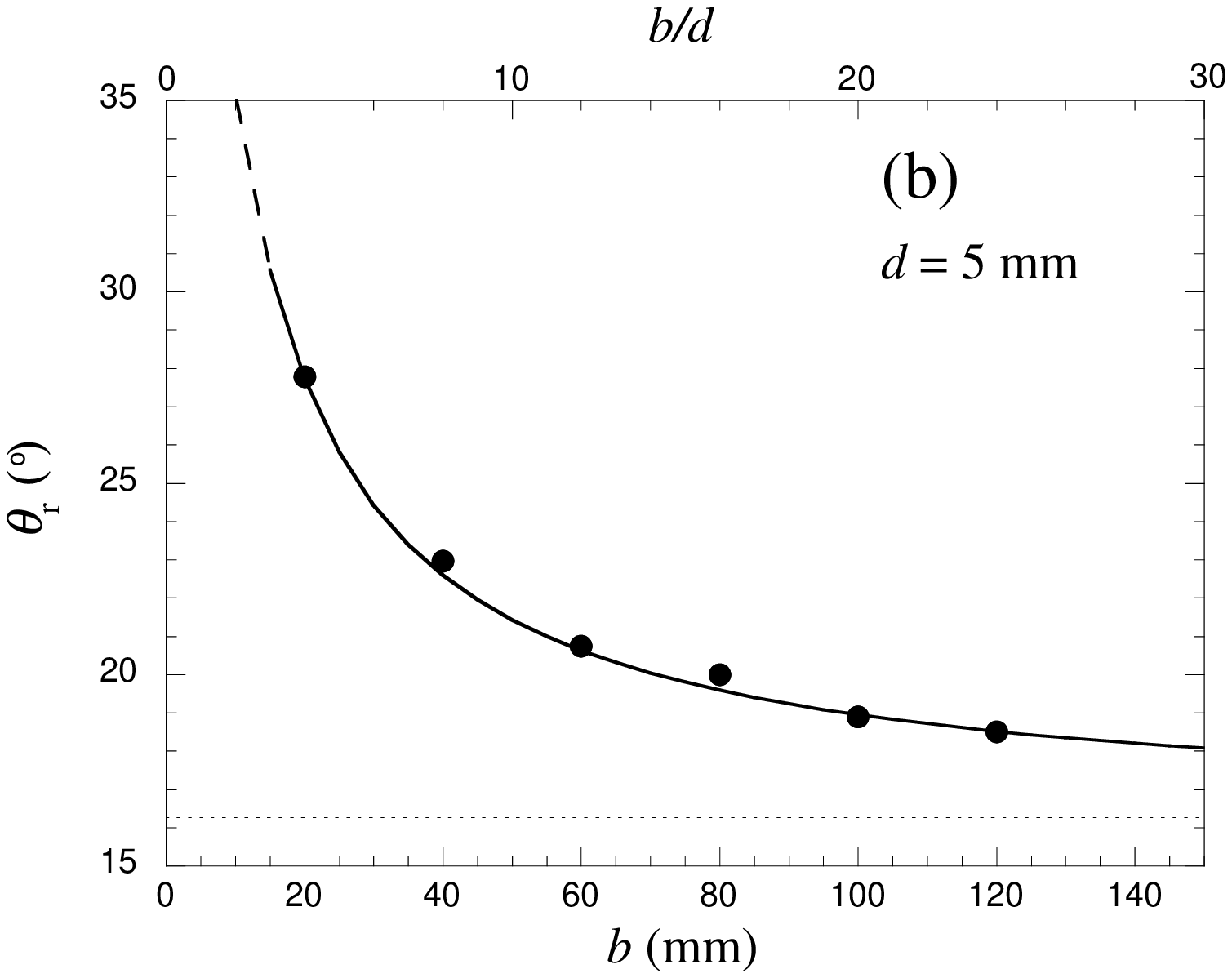,width=0.9\linewidth}}
    \end{minipage}
    
	\caption{Pile angles $\theta_{\rm{m,r}}$ as a function of the gap width $b$. (a) Our experimental
	data for $\theta_{\rm{m}}$ ($\blacktriangle$) and $\theta_{\rm{r}}$ 
	($\bullet$) for 
	$d = 3 \, \rm{mm}$ glass beads in water (error bars
	correspond to the standard deviation). (b) Numerical results of 
	\cite{zhou} for
	$d = 5 \, \rm{mm}$ glass beads in vacuum. Solid lines correspond to our model 
	eq. \ref{eq6} with (a) ($\theta_{\rm{m}}^{\infty}= 24.75^{\circ}, \, B_{\rm{m}} = 6 \, \rm{mm}$) 
	and ($\theta_{\rm{r}}^{\infty} = 23^{\circ}, \, B_{\rm{r}} = 4 \, \rm{mm}$) and
	(b) ($\theta_{\rm{r}}^{\infty} = 16.25^{\circ}, \, B_{\rm{r}} = 9.8 \, 
	\rm{mm}$). Horizontal dotted lines correspond 
	to the asymptotic values $\theta_{\rm{r}}^{\infty}$ and 
	$\theta_{\rm{m}}^{\infty}$ .}
	\label{fig3}
\end{figure}

Figure 3a shows typical variations of both the angle of movement 
$\theta_{\rm{m}}$ and the angle of repose $\theta_{\rm{r}}$ as a function of 
the gap width $b$. Increasing the gap makes $\theta_{\rm{m}}$ and 
$\theta_{\rm{r}}$ decrease towards respective constant values 
$\theta_{\rm{m}}^{\infty}$ and $\theta_{\rm{r}}^{\infty}$. 
One can also notice that the avalanche amplitude 
($\theta_{\rm{m}}-\theta_{\rm{r}}$) increases with a 
decreasing gap width as it has been previously reported \cite{evesque}. We propose now a simple physical model taking into account the lateral 
wall effects on pile angles.\\
The presence of confining walls is well known to play a key role for the 
saturation of the normal stresses with the depth of a granular pile. Let us 
first look at this Janssen effect in our configuration. In a pile making an 
angle $\theta$ to the horizontal, let us consider a thin layer of 
thickness $dh$, length $dl$ and width $b$ (corresponding to the 
whole gap) located at the depth $h$ under the pile surface (fig. 4). 
\begin{figure}[h!]
    
    \centerline{\epsfig{file=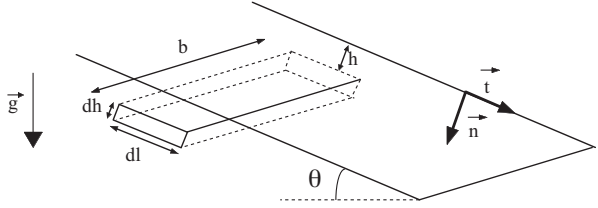,width=0.55\linewidth}}
     \caption{Scheme of the pile described as a continuous medium.}
     \label{fig4}
 \end{figure}
At equilibrium, the $\vec{n}$ component (normal to the pile 
interface) of the weight of the layer is balanced by the gradient of 
the stress $\sigma_{nn}$ (here called the pressure $p$) and by friction on 
the walls. Following the Janssen analysis, a part of the pressure 
(that is supposed to depend only on $h$) is redistributed normal to each wall through the Janssen coefficient $K$ which induces a fully mobilised friction 
force with a friction coefficient $\mu$ so that the equilibrium equation along the $\vec{n}$ direction writes
\begin{equation}
\label{eq2}
\rho g \cos\theta \, b \, dh \, dl +[p(h)-p(h+dh)] \,b \,dl - 2 \, K \, 
\mu  \, p(h) \, dh \, dl=0,
\end{equation}
where $\rho$ is the mean pile density and $g$ is the gravity 
acceleration. By integrating eq. (\ref{eq2}) we obtain the pressure expression
\begin{equation}
\label{eq3}
p(h)=\rho \, g \, cos\theta \, \frac{b}{2 \, K \mu} \left[ 1-\exp 
\left( -\frac{2\, K \mu h}{b}  \right) \right].
\end{equation}
A surface avalanche starts when the equilibrium along the $\vec{t}$ direction (tangent to the pile interface) is broken for a layer of 
depth $h_{\rm{crack}}$.
Without lateral walls, this arises when the pile reaches the critical 
angle $\theta_{\rm{m}}^{\infty}$ to the horizontal, {\it{i.e.}} when 
the $\vec{t}$ component of 
the layer weight reaches the critical value
\begin{equation}
\label{eq4}
F^{\infty}=\rho \, g \, b \, h_{\rm{crack}} \, dl \, \sin \theta_{\rm{m}}^{\infty}.
\end{equation}
When lateral walls are present, the heap stability increases because of the friction 
force on the walls along the $\vec{t}$ direction. Thus, when the lateral 
walls are $b$ apart, we assume that the 
avalanche starts at the same $h_{\rm{crack}}$ for a larger critical pile angle 
$\theta_{\rm{m}}(b)$ given by
\begin{equation}
\label{eq5}
\rho \, g \, b \, h_{\rm{crack}} \, dl \, \sin \theta_{\rm{m}}(b)=F^{\infty}+2 \, K 
\, \mu \,dl \, \int_{0}^{h_{\rm{crack}}} p(h) \, dh.
\end{equation}
By introducing the pressure expression (eq. \ref{eq3}) and the 
$F^{\infty}$ expression
(eq. \ref{eq4}) in eq. (\ref{eq5}) we obtain the following equation that relates the 
movement angle $\theta_{\rm{m}}$ to the gap width $b$:
\begin{equation}
\label{eq6}
\frac{\sin \theta_{\rm{m}}(b)-\sin \theta_{\rm{m}}^{\infty}}{\cos 
\theta_{\rm{m}}(b)}=1-\frac{b}{B_{\rm{m}}} \left[ 1- \exp \left( 
-\frac{B_{\rm{m}}}{b} \right) \right],
\end{equation}
which reduces for $b\gg B_{\rm{m}}$ to
\begin{equation}
\label{eq7}
\frac{\sin \theta_{\rm{m}}(b)-\sin \theta_{\rm{m}}^{\infty}}{\cos 
\theta_{\rm{m}}(b)} \simeq \frac{B_{\rm{m}}}{2b},
\end{equation}
where $B_{\rm{m}}$ is the characteristic length of lateral wall 
effects given by
\begin{equation}
\label{eq8}
B_{\rm{m}}=2 \, K \, \mu \, h_{\rm{crack}}.
\end{equation}
Equation (\ref{eq6}) involves the two physical parameters 
$\theta_{\rm{m}}^{\infty}$ and $B_{\rm{m}}$. Its predicts 
that $\theta_{\rm{m}}$ decreases when $b$ increases and reaches $\theta_{\rm{m}}^{\infty}$ when $b$ tends towards infinity. 
For the typical value  $\theta_{\rm{m}}^{\infty}= {\rm{25}}^{\circ}$, the 
maximal value $\theta_{\rm{m}}^{0}$ that would be obtained for $b = 0$ 
is equal to ${\rm{62}}^{\circ}$. It is worth noting that the 
$\theta_{\rm{m}}$ decrease does not follow a classical 
exponential decay 
(here $(\theta_{\rm{m}}(b)-\theta_{\rm{m}}^{\infty})/(\theta_{\rm{m}}^{0}-\theta_{\rm{m}}^{\infty})=1/e$ for $b\approx 2 B_{\rm{m}}$).
Indeed, $\theta_{\rm{m}}$ tends more slowly towards its asymptotic 
value $\theta_{\rm{m}}^{\infty}$.
In the Janssen analysis, the granular medium is considered as a 
continuum. Already questionable considering a static pile, this hypothesis is all 
the more questionable considering a flowing layer. Nevertheless, the Janssen model 
enables to account for dynamic effects, such as the constant flow of 
hourglasses. Assuming that eq. (\ref{eq6}) can also be applied just before 
an avalanche stops, one obtains the same expression for the angle of repose 
$\theta_{\rm{r}}(b)$ whose asymptotic value is $\theta_{\rm{r}}^{\infty}$ 
and characteristic length is $B_{\rm{r}}=2 \, K \, \mu \, 
h_{\rm{freeze}}$, $h_{\rm{freeze}}$ being the flowing layer height when the flow 
is about to stop. Note that both $K$ and $\mu$ coefficients could be slightly 
different in $B_{\rm{m}}$ and $B_{\rm{r}}$ expressions as they correspond either to a static 
or dynamic case.\\
In fig. 3 the two typical data sets from our experiments (a) and from the numerical simulations of 
Zhou et {\it{al.}} \cite{zhou} are fitted with our model eq. (\ref{eq6}) . 
These curves illustrate the good agreement we obtain between our model and all the data sets. The exponential decay used in ref. 
\cite{gras,bolt,zhou} fits as well all the data but it involves 
three  fitting parameters, and is not based on any physical argument. 
Note that our continuum model is only pertinent for $b/d \, \gtrsim 
\, 3$ (suggested by the dashed lines in fig. 3) as the stress redistribution needs a few beads to be effective. 
In this range, eq. (\ref{eq7}) is a very good approximation of the 
full eq. (\ref{eq6}).
\begin{figure}[h!]
    
     \begin{minipage}[b]{0.47\linewidth}
	\centerline{\epsfig{file=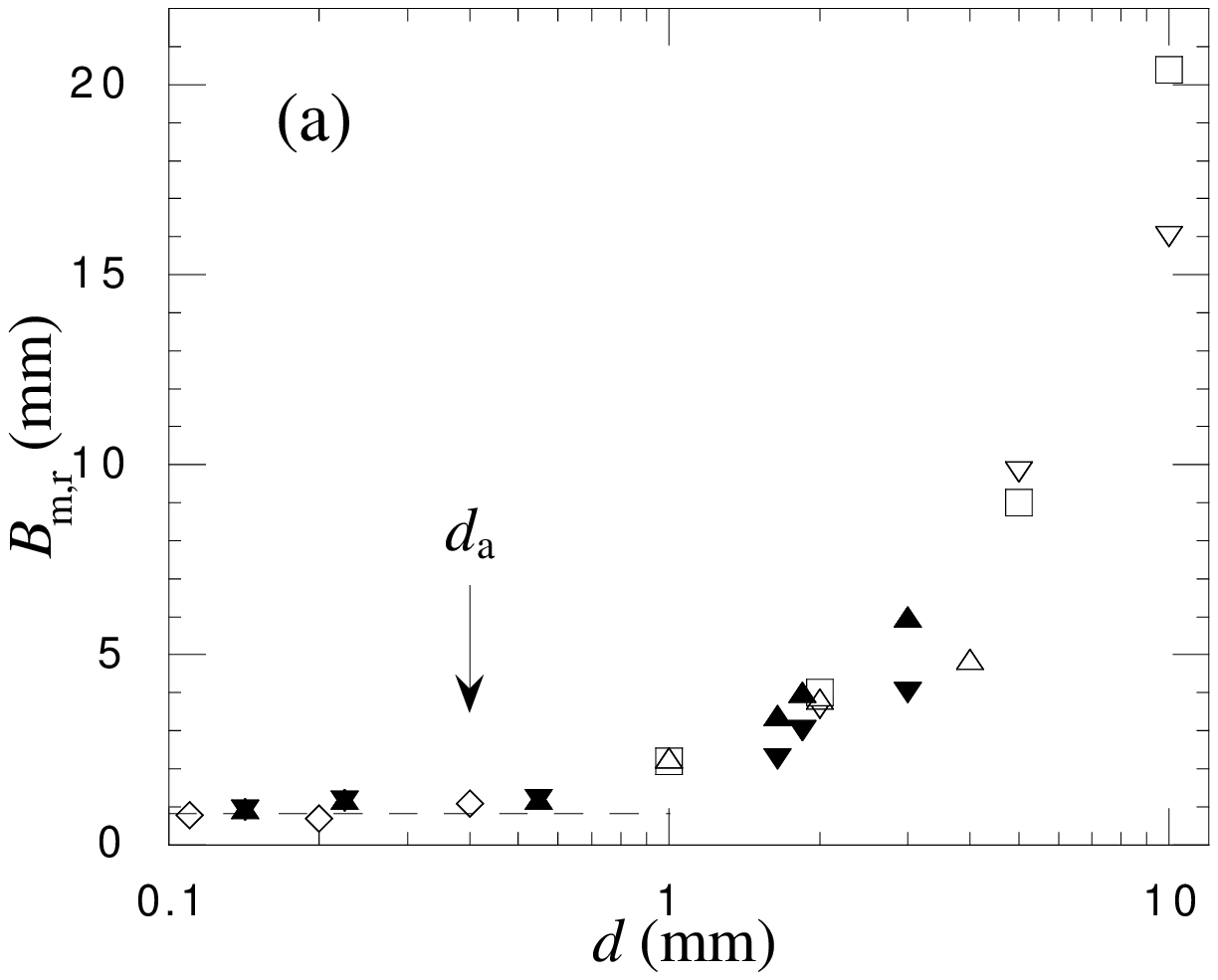,width=0.9\linewidth}}
     \end{minipage}
     \hfill
    \begin{minipage}[b]{0.47\linewidth}
	\centerline{\epsfig{file=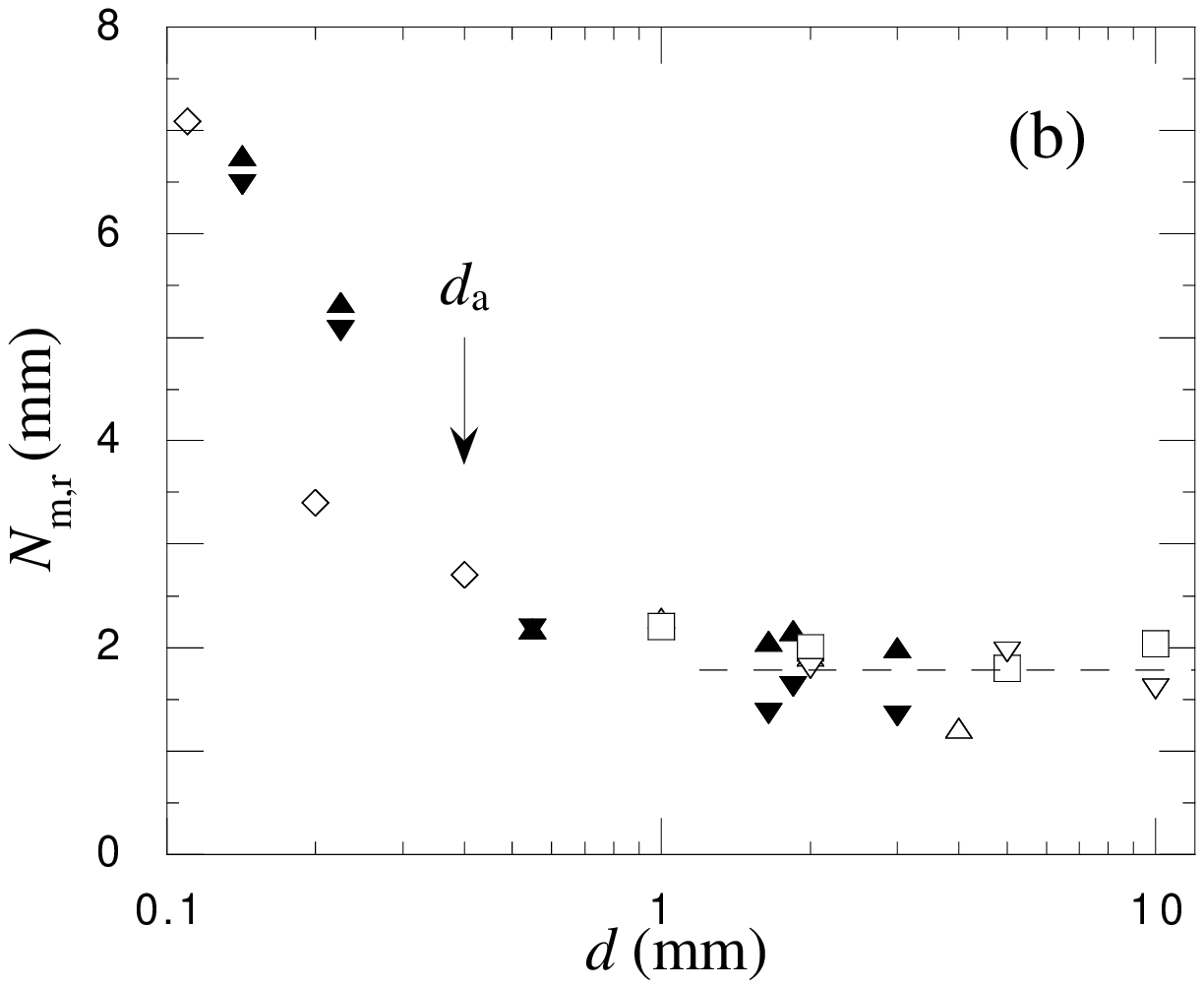,width=0.9\linewidth}}
    \end{minipage}
    
	\caption{Characteristic range of wall effects $B_{\rm{m,r}}$ (a) and 
	$N_{\rm{m,r}} = B_{\rm{m,r}}/d$ (b) as a function of the bead diameter $d$ extracted 
	from our experimental data for $\theta_{\rm{m}}$ ($\blacktriangle$)
	and $\theta_{\rm{r}}$ ($\blacktriangledown$) in water and from experimental data 
	of ref.\cite{gras} for $\theta_{r}$ in air ($\diamond$),
	ref. \cite{bolt} for $\theta_{m}$ in air ($\vartriangle$),
	and ref. \cite{zhou} for $\theta_{\rm{r}}$ in air ($\square$), 
	and from the numerical data of ref. \cite{zhou} for $\theta_{\rm{r}}$ in vacuum ($\triangledown$).}
	\label{fig5}
\end{figure}

In order to extract the evolution of the characteristic lengths $B_{\rm{m,r}}$ with the bead diameter $d$, 
our experimental data but also data from the three former studies mentioned 
above \cite{gras,bolt,zhou} have been analysed in the light of this model. For our experiments 
the asymptotic angle values are all found between $22.5^{\circ}$ and $25.5^{\circ}$ for $\theta_{\rm{m}}^{\infty}$ and between 
$22^{\circ}$ and $24.5^{\circ}$ for $\theta_{\rm{r}}^{\infty}$. Figures 5a and 
5b show the characteristic length $B_{\rm{m,r}}$ and the 
corresponding characteristic number of beads $N_{\rm{m,r}}=B_{\rm{m,r}}/d$ as a function of the bead 
diameter $d$ for our experiments in water and other results in air \cite{gras,bolt,zhou}. Considering 
this, all data collapse remarkably well onto a single curve. This suggests that neither
the Janssen coefficient $K$ nor the friction coefficient $\mu$ for glass beads in air or in
water, with glass or Plexiglas lateral walls, change much from one experimental set-up to another
\cite{bolt-rug}. However, according to the different ways of making a heap (in water or in air, with a 
rotating drum or by the discharging method, etc), the asymptotic values 
$\theta_{\rm{m,r}}^{\infty}$ vary ({\it e.g.},   
varies from $14^{\circ}$ to $26^{\circ}$).\\
Figures 5a and 5b clearly put in light two regimes depending on 
the bead diameter. For large beads ($d > 1 \, \rm{mm}$), $N_{\rm{m,r}}$ is found constant 
($N_{\rm{m,r}} \approx 1.8 \pm 0.4$), 
leading to a characteristic length $B_{\rm{m,r}}$ proportional to $d$. For small 
beads ($d < 1\, \rm{mm}$), $N_{\rm{m,r}}$ 
strongly increases with decreasing bead diameters, leading to a constant characteristic 
length ($B_{\rm{m,r}} \approx 0.8  \pm 0.2 \, \rm{mm}$) as it was previously observed 
in air \cite{gras}.\\
Let us first focus on large beads. In that case the relevant parameter that governs
wall effects on pile angles is then the number $N_{\rm{m,r}}$ of beads in the gap width, as $N_{\rm{m,r}}$ 
is constant. $N_{\rm{m,r}}$ being of the order of two beads (cf. fig. 5b), 
63\% of wall effects on pile 
angles have disappeared for a gap width $b \approx 4 \, d$. This constant number of beads, which can be 
interpreted as the consistent length of a force network, can be understood seeing the stress
redistribution to the lateral walls acting geometrically through contact networks. One may notice that such 
a characteristic length, of the order of a few bead diameters, rises from different experiments 
~\cite{long-char}. Moreover, regarding eq. (\ref{eq8}) and considering 
the typical value $K \mu \, \sim 0.2$ \cite{jans}, 
$h_{\rm{crack}}$ and $h_{\rm{freeze}}$ are found roughly equal to 4 - 5 bead diameters, which is consistent with 
the thickness of the granular flowing layer we observed.\\
If we consider our results, $N_{\rm{r}}$ is slightly smaller than 
$N_{\rm{m}}$. 
Apart from the fact that $h_{\rm{crack}}$ and $h_{\rm{freeze}}$ can possibly be different,
this difference may be explained by a lower value 
of the friction coefficient $\mu$ and/or of the Janssen coefficient 
$K$ in the dynamical case ~\cite{dilat}.\\
Let us look now at the small bead regime (typically $d < 1 \, \rm{mm}$). For small 
beads the characteristic range of wall effects $B_{\rm{m,r}}$ is no more proportional 
to $d$ but tends, for vanishing bead diameters, to a constant value equal 
to roughly $0.8 \, \rm{mm}$, regarding 
experiments performed both in air and in water (fig. 5a). One possible explanation 
for this phenomenom is 
that sub-millimetric beads aggregate. Cohesion due to capillary bridges between grains has been 
evoked as a possible explanation for aggregate formation in air 
\cite{gras}. However, capillary bridges 
can not form in our fully immersed packing. Moreover, the accordance between our experiments in 
water and experiments in air suggests to look for a unique cause that explains aggregates. We 
show here that surface forces such as van der Waals forces between beads lead to aggregates and 
can explain the constant value of $B_{\rm{m,r}}$.\\
The van der Waals energy between two spheres of respective radius 
$R_{\rm{1}}$ and $R_{\rm{2}}$, a distance $\delta$ apart is 
$W(\delta)=-A \, R_{\rm{1}} \, R_{\rm{2}} / 6 \, \delta \, ( 
R_{1}+R_{2})$,
where $A$ is the Hamaker constant \cite{israelach}. We can estimate the aggregate size so that a bead belongs to the aggregate
if the resulting van der Waals forces are larger than the whole aggregate weight.
Our glass beads (commercial crushing glass beads) are rough with a roughness size
that is independent of the bead size and that is of the order of one micron. Then, 
van der Waals forces between two beads of radius $R_{\rm{1}}$ in contact may be calculated between a
bead and a roughness of radius $R_{\rm{2}}$ ($R_{\rm{2}} \ll R_{\rm{1}}$) at a distance 
of perfect contact, {\it i.e.} corresponding
to the molecular size $\delta$ \cite{calcVdW}. Considering three contacts 
for a peripheral 
bead with the rest of the aggregate, the aggregate diameter is 
\begin{equation}
\label{diameter}
d_{\rm{a}}= \left( \frac{3 \, A \, R_{\rm{2}}}{\pi \, c_{0} \, 
\delta^{2} \, \Delta \rho \, g} \right)^{1/3},
\end{equation}
where $c_{0}$ is the packing fraction and $\Delta \rho$ the apparent bead density. 
The aggregate diameter, thus found independent of the particle size, is $d_{\rm{a}} \approx 0.3 \, \rm{mm}$ 
for glass beads immersed in water and $d_{\rm{a}} \approx 0.5 \, \rm{mm}$ for glass beads in dry air 
(with $A = 10^{-20} \, \rm{J}$ and $\Delta \rho = 1500 \, \rm{kg.m^{-3}}$ in the water case, $A = 
10^{-19} \, \rm{J}$ and $\Delta \rho = 2500 \, \rm{kg.m^{-3}}$ in the dry air 
case, $\delta  = 0.2 \, \rm{nm}$
, $R_{\rm{2}} = 1 \, \rm{\mu m}$, $c_{0} = 0.6$ and $g = 9.8\, \rm{m.s^{-2}}$). One observes on fig. 5 
that the value $d_{\rm{a}} \approx 0.4 \, \rm{mm}$ predicts well the transition between the two regimes. Furthermore, 
as well as $B_{\rm{m,r}} \approx 2 d$ for large beads ($d > d_{\rm{a}}$) 
$B_{\rm{m,r}} \approx 
2 d_{\rm{a}} \approx 0.8 \, \rm{mm}$ for small beads ($d < 
d_{\rm{a}}$).\\ 
Thus, the characteristic length of wall effects $B_{\rm{m,r}}$ always corresponds 
to a constant number of either beads (for $d > d_{\rm{a}}$) or 
aggregates (for $d < d_{\rm{a}}$).\\

The presence of close lateral walls increases the heap stability in the same way in air 
and in water. This phenomenon can be explained by particle arches 
that direct a part of the weight to the walls, thus inducing friction. Our physical model based 
on arching effects is in accordance with our experimental results and results of 
former studies. For non-cohesive spheres the characteristic length of wall effects on pile angles 
is found to be proportional to the bead diameter $d$. 
By contrast, due 
to van der Waals forces, small beads ({\it{i.e.}} for glass, $d < 0.4 \, 
\rm{mm}$) aggregate either in air or in water which makes the characteristic 
length be constant and no more proportional to the bead diameter.
\acknowledgments
The authors thank N. Merakchi for his help in preliminary 
experiments and O. Dauchot for stimulating discussions, and are grateful to Y. C. Zhou and A. B. Yu as well as Y. Grasselli 
who have kindly communicated their data. The ACI ``Jeunes Chercheurs" 
$ \mbox{n}^{\circ}$ 
2178 and ``Pr\'{e}vention 
des Catastrophes Naturelles" of the French Ministry of Research have supported this work.


\begin{thebibliography}{99}

\bibitem {mouv} 
\Name {Bagnold R. A.}  
\REVIEW {Proc. Roy. Soc. A}{225}{1954}{49}.

\bibitem {divers} 
\Name {Van Burkalow A.}  
\REVIEW {Bull. of the Geological Society of America}{56}{1945}{669}.

\bibitem {cohesion} 
\Name {Bocquet L. {\it et al.}}  
\REVIEW {Nature}{396}{1998}{735}.

\bibitem {liu} 
\Name {Liu C., Jaeger H. M. \and Nagel S. R.}  
\REVIEW {Phys. Rev. A}{43}{1991}{7092}.

\bibitem {gras} 
\Name {Grasselli Y. \and Herrmann H. J.}  
\REVIEW {Physica A}{246}{1997}{301}.

\bibitem {bolt} 
\Name {Boltenhagen P.}  
\REVIEW {Eur. Phys. J. B}{12}{1999}{75}.

\bibitem {zhou} 
\Name {Zhou Y. C., Xu B. H. \and Yu A. B.}  
\REVIEW {Phys. Rev. E}{64}{2001}{021301};
\Name {Zhou Y. C. {\it et al.}}  
\REVIEW {Powder Technol.}{125}{2002}{45}.

\bibitem {jans} 
\Name {Janssen H. A.}  
\REVIEW {Z. Vereins Deutsh Ing.}{39}{1895}{1045};
\Name{Duran J.}
\Book{Sand, Powders and grains}
\Publ{Springer, New-york, 2000}.


\bibitem {intermit} 
\Name {Rajchenbach J.}  
\REVIEW {Phys. Rev. Lett.}{65}{1990}{2221}.

\bibitem {fauve} 
\Name {Caponeri M. {\it et al.}}
\Book {{\rm in} Mobile Particulate systems}
\Publ {Kluwer, Dordrecht, 1994}{ 331};
\Name {Dury C. M. {\it et al.}}
\REVIEW {Phys. Rev. E}{57}{1998}{4491}.

\bibitem {evesque} 
\Name {Evesque P.}  
\REVIEW {Phys. Rev. A}{43}{1991}{2720}.

\bibitem {bolt-rug}
The characteristic length of wall effect increases when increasing significantly 
the roughness of the lateral walls \cite{bolt}.  This may be explained by a larger 
friction coefficient $\mu$ in eq. (\ref{eq8}).

\bibitem {long-char}
For granular flows on rough inclined planes, 
Pouliquen [\REVIEW {Physics of Fluids}{11}{1999}{542}] and 
Daerr and Douady [\REVIEW {Nature}{399}{1999}{241}] show 
that the angle of repose quickly decreases with an 
increasing critical material thickness to a constant 
value with a characteristic length of a few bead diameters. In Couette experiments, 
velocity profiles in the shear layer of a granular 
media [\Name {Mueth D. M. {\it et al.}} \REVIEW {Nature}{406}{2000}{385}; 
\Name {Bocquet L. {\it et al.}} \REVIEW {Phys. Rev. E}{65}{2001}{011307}],
as well as the mean velocity 
of the creep motion of granular surface flow [\Name {Komatsu T. S.} 
\REVIEW {Phys. Rev. Lett.}{86}{2001}{1757}] 
decay exponentially with characteristic lengths of a few bead diameters. 

\bibitem {dilat}
As granular media dilate under shear, the packing 
fraction of the flowing layer is lower than the one for 
a static pile. Thus, stress redistribution to the walls 
may be less efficient with a smaller $K$ value.

\bibitem {israelach}
\Name {Israelachvili J.}
\Book {Intermolecular and Surface Forces}
\Publ {Academic Press, 1991}.


\bibitem {calcVdW}
\Name {Albert R. {\it et al.}}
\REVIEW {Phys. Rev. E}{56}{1997}{R6271}. 




\end{thebibliography}
\end{document}